\documentclass[preprint, cite]{epl}
\usepackage{graphicx}
\title{Equation of motion of the triple contact line along an inhomogeneous surface}
\shorttitle{Triple contact line motion}
\author{Vadim S. Nikolayev\thanks{E-mail:
\email{vnikolayev@cea.fr}} \and Daniel A. Beysens} \shortauthor{V. S. Nikolayev \etal} \institute{ESEME,
Service des Basses Temp\'eratures, DSM/DRFMC, CEA-Grenoble, France\thanks{Mailing address: CEA-ESEME,
Institut de Chimie de la Mati\`ere Condens\'{e}e de Bordeaux,
  87, Avenue du Dr. Schweitzer, 33608 Pessac Cedex, France}}

\begin{document}
\date\today
\maketitle
\begin{abstract}
The wetting flows are controlled by the contact line motion. We derive an equation that describes the slow
time evolution of the triple solid-liquid-fluid contact line for an arbitrary distribution of defects on a
solid surface. The capillary rise along a partially wetted infinite vertical wall is considered. The contact
line is assumed to be only slightly deformed by the defects. The derived equation is solved exactly for a
simple example of a single defect.
\end{abstract}

\section{Introduction}
The wetting flows (where the triple solid-liquid-fluid contact line is present) are important for many
practical applications ranging from the metal coating to the medical treatment of the lung airways. The
contact line statics and dynamics attracted a lot of attention from the scientific community during the last
decades. It became clear that the hydrodynamics in the region of the liquid wedge close to the contact line
(which we will call CLR - Contact Line Region) differs from the hydrodynamics in the bulk of the liquid.
Because of the contact line singularity \cite{DeGennes}, the fluid motion in presence of the contact line
appears to be much slower than without it. This difference can be accounted for by the introduction of the
anomalously large energy dissipation inside the CLR \cite{DeGennes}.

For practical purposes, one needs to know the dynamics of the liquid surface influenced by this dissipation.
This influence is especially strong in the very common case of (i) low viscosity fluids that (ii) wet
partially the solid with (iii) no precursor film on it. In this case the bulk dissipation is particularly
small with respect to the large CLR dissipation \cite{JFM}. The contact line motion is very slow and the
liquid surface can be described in the quasi-static approximation \cite{YP}. The dissipation in the liquid
(energy per unit time) can then be approximated by the dissipation in the CLR as \cite{DeCon}
\begin{equation}\label{diss}
\int{\xi \, v_n^2\over 2}\;{\rm d}l,
\end{equation}
where $v_n$ is the normal component of the contact line velocity and the integration is performed along the
contact line. The generalized dissipation coefficient $\xi$ is a constant that is assumed to be much larger
that the liquid shear viscosity. The expression (\ref{diss}) assumes that the contribution of  a piece of the
CLR to the dissipation is proportional to the contact line length. Then the $v_n^2$ term is leading for small
$v_n$. The dissipation coefficient $\xi$ can be obtained e.g. from measurements of the kinetics of relaxation
of an oval sessile drop toward its equilibrium shape (see \cite{JFM}, where it was found $10^{7}$ times
larger than the shear viscosity). It can also be obtained from the measurements of $v_n$ and the dynamic
contact angle $\theta$ by using the expression
\begin{equation}
v_n = {\sigma\over\xi}\,(\cos\theta_{eq}-\cos\theta),  \label{cos}
\end{equation}
where $\theta_{eq}$ is the equilibrium value of the contact angle and $\sigma$ is the surface tension.
Eq.~\ref{cos} is common for many contact line motion models. It can be shown \cite{DeCon} that Eq.~\ref{cos}
follows from the expression (\ref{diss}) for the cases where $v_n$ does not vary along the contact line (i.e.
for the contact line of constant curvature).

Usually, one is interested to know the displacement of the contact line (or its statistical properties in the
case of the irregular solid) because its position serves as a boundary condition for the determination of the
shape of the liquid surface. The recent articles on this subject show that no general approach to this
problem is accepted. It is recognized generally \cite{PomVan,Mezard,Vannim} that the static contact line
equation should be non-local because the contact line displacement at one point influences its position at
other points through the surface tension. In dynamics, the local relations similar to Eq.~\ref{cos} were
considered universal for a long time.  In our previous work \cite{Relax} we developed a non-local dynamic
approach. We showed that if non-locality is taken into account, Eq.~\ref{cos} is not valid in the general
case where $v_n$ varies along the contact line. This variation can be due either to the initially
inhomogeneous contact line curvature (like in \cite{Relax}) or to a inhomogeneous substrate. In this Letter
we analyze this latter case. We use this non-local approach to derive an equation of motion for the very
common case of the capillary rise of the liquid along a vertical solid wall with the account of the surface
defects.

\section{Derivation of the equation}

\begin{figure}[tbh]
  \begin{center}
  \includegraphics[height=4cm]{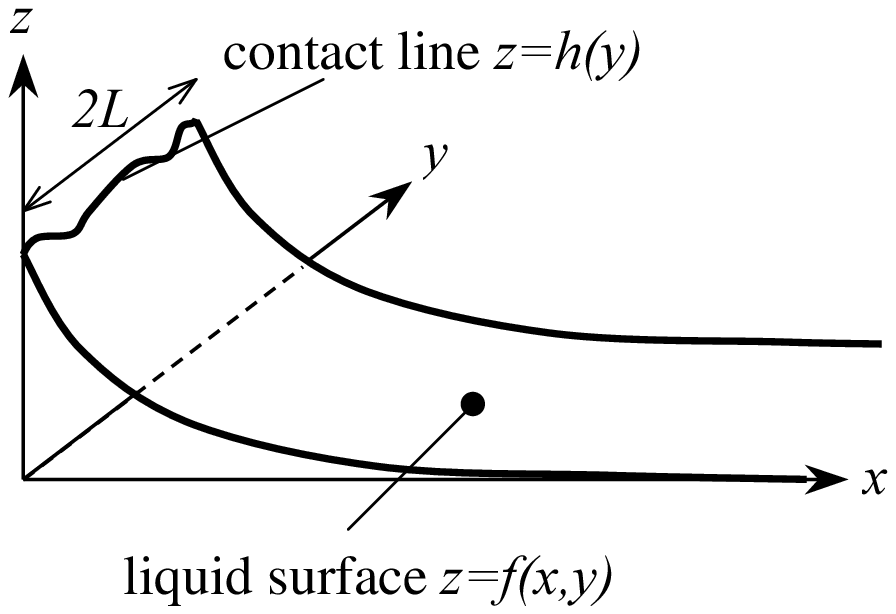}
  \end{center}
\caption{ Reference system to describe the shape of the liquid surface.} \label{Wilh}
\end{figure}

A Cartesian reference system $xyz$ is chosen in such a way that the liquid surface far from the wall (defined
by the $y-z$ plane) coincides with the $x-y$ plane, see Fig.~\ref{Wilh}. The liquid surface described by the
equation $z=f(x,y)$ is assumed to be weakly deformed so that
\begin{equation}\label{weak}
	|\partial f/\partial x|,|\partial f/\partial y|\ll 1.
\end{equation}
The contact line can be described by the equation $z=h(y)\equiv f(0,y)$. A piece of the liquid surface of
length $2L$ in the $y$-direction (see Fig.~\ref{Wilh}) is considered so that the final form for the equation
of motion is found in the limit $L\rightarrow\infty$.

First, following the approach \cite{PomVan}, we find the energy $U$ of the liquid (per contact line length)
as a functional of $h$. It is convenient to break $U$ into two terms, $U=U_1+U_2$. In the approximation
(\ref{weak}), the first term reads
\begin{equation}\label{U1}
U_1={\sigma\over 4L}\int_{-L}^L \textrm{d}y\int_0^\infty \textrm{d}x\left[ \left({\partial f\over\partial
x}\right)^2+\left({\partial f\over\partial y}\right)^2+ {f^2\over l_c^2} \right],
\end{equation}
where $l_c=[\sigma/(\rho g)]^{1/2}$ is the capillary length, $\rho$ is the liquid density and $g$ is the
gravity acceleration. The defects are accounted for in the second term \cite{PomVan,Garoff}
\begin{equation}\label{U2}
U_2=-{\sigma\over 2L}\int_{-L}^L \textrm{d}y\int_0^{h(y)} c(y, z)\,\textrm{d}z
\end{equation}
by the fluctuations of the function $c(y,z)$ which is the difference of the surface energies (in $\sigma$
units) of the gas-solid and liquid-solid interfaces at the point $(y,z)$ of the solid surface. According to
the Young expression, $c(y,z)=\cos[\theta_{eq}(y,z)]$, where $\theta_{eq}$ is the local value of the
equilibrium contact angle. When the surface defects exist, $\theta_{eq}$ is an arbitrary function of $y,z$.

In the quasi-equilibrium approximation, the surface shape $f$ is found by minimization of the functional $U$
resulting in the equation
\begin{equation}\label{feq}
(\partial^2 f/\partial x^2)+(\partial^2 f/\partial y^2)=f/ l_c^2,
\end{equation}
which can be solved by separating the variables by using the boundary condition $f(x\rightarrow\infty)=0$ and
assuming that the function $f(x,y)$ is bounded at $y\rightarrow\pm\infty$. The solution reads
\begin{equation}\label{f}
f=h_0\exp(-x/l_c)+\sum_{n=1}^\infty[a_n\cos(\pi ny/L)+ b_n\sin(\pi ny/L)]\exp\left(-x\sqrt{l_c^{-2}+\pi^2
n^2/L^2}\right),
\end{equation}
where the coefficients $h_0$, $a_n$, $b_n$ are the coefficients for the Fourier series
$$h(y)=f(x=0)=h_0+\sum_{n=1}^\infty[a_n\cos(\pi ny/L)+ b_n\sin(\pi ny/L)].$$
They can thus be related to $h(y)$ by the expressions
\begin{equation}\label{abh}
 a_n={1\over L}\int_{-L}^L h(y)\cos{\pi ny\over L}\,\textrm{d}y,\quad
 b_n={1\over L}\int_{-L}^L h(y)\sin{\pi ny\over L}\,\textrm{d}y,\quad h_0={1\over 2L}\int_{-L}^L
h(y)\,\textrm{d}y.
\end{equation}
One can notice that $h_0$ is simply the horizontally averaged displacement of the contact line. The back
substitution of Eq.~(\ref{f}) with $h_0$, $a_n$, $b_n$ replaced by their values (\ref{abh}) into
Eq.~(\ref{U1}) results in the functional
\begin{equation}\label{funcU1}
U_1={\sigma\over 4L^2}\int_{-L}^L \textrm{d}y\int_{-L}^L \textrm{d}y'h(y)h(y')\Biggl[(2l_c)^{-1} +
\sum_{n=1}^\infty\cos{\pi n(y-y')\over L}\sqrt{{1\over l_c^2}+{\pi^2 n^2\over L^2}}\;\Biggr].
\end{equation}
Eq.~(\ref{U2}) together with Eq.~(\ref{funcU1}) define the functional $U[h(y)]$.

Now, one needs to calculate the dissipation function $T$ (per unit length). In the approximation
(\ref{weak}), $v_n^2{\rm d}l\approx \dot h^2{\rm d}y$ in Eq.~(\ref{diss}) and
\begin{equation}\label{T}
T={1\over 2L}\int_{-L}^L{\xi \, \dot h^2\over 2}\;{\rm d}y,
\end{equation}
where the dot means a time derivative. The equation of motion can be written in the quasi-static
approximation as
\begin{equation}\label{var}
  {\delta U[h]\over \delta h}=-{\delta T[\dot h]\over \delta \dot h},
\end{equation}
where $\delta\ldots\over\delta\ldots$ means variational derivative. By substituting
Eqs.~(\ref{U2},\ref{funcU1},\ref{T}) into Eq.~(\ref{var}) and taking the variational derivatives, one obtains
the dynamic equation of the contact line motion:
\begin{equation}\label{hL}
 \dot h(y)=-{\sigma\over\xi} \Biggl[h_0/l_c-c(y, h)+
 {1\over L}\sum_{n=1}^\infty\int_{-L}^L
\textrm{d}y'h(y')\cos{\pi n(y-y')\over L}\sqrt{l_c^{-2}+{\pi^2 n^2\over L^2}}\Biggr],
\end{equation}
where $h$ and $h_0$ are assumed to be time dependent. The final form for the integral equation for the
contact line motion can be obtained by taking the limit $L\rightarrow\infty$:
\begin{equation}\label{hh}
 \dot h(y)={\sigma\over\xi} \biggl\{c(y, h(y))-
 {1\over\pi}\int_0^\infty\textrm{d}p \int_{-\infty}^\infty
\textrm{d}y'h(y')\cos[p(y-y')]\sqrt{l_c^{-2}+p^2}\biggr\}.
\end{equation}

To treat the contact line equation, it is convenient to introduce the spatial fluctuation $h_1=h-h_0$ and
solve the equations for $h_1$ and $h_0$ separately. First, we derive the dynamic equation for $h_0$ by
integrating Eq.~(\ref{hL}) over $y$ from $-L$ to $L$, changing the integration order and dividing by $2L$.
The integrals of all terms in the sum are zero and one obtains the following equation for $h_0$

\begin{equation}\label{h0}
\dot h_0=-{\sigma\over\xi} (h_0/l_c-c_0),\quad c_0={1\over 2L}\int_{-L}^L c(y, h_1+h_0)\,\textrm{d}y,
\end{equation}
where $c_0$ is the (time dependent) horizontal average of $c$. By subtracting Eq.~(\ref{h0}) from
Eq.~(\ref{hh}) one obtains the equation for $h_1$ which has the same form as Eq.~(\ref{hh}) where $c$ should
be replaced by its fluctuation $c_1=c(y, h_1+h_0)-c_0$. Equation  for $h_1$ simplifies in the Fourier space.
By making use of the convolution theorem \cite{Korn}, one obtains
\begin{equation}\label{h1F}
  \dot {\tilde h}_1={\sigma\over\xi} \left(\tilde c_1-\tilde h_1\sqrt{l_c^{-2}+k^2}\right),
\end{equation}
where the tilde over a variable denotes its Fourier transform with the parameter $k$,
e.g.
\begin{equation}\label{ft}
  \tilde h_1=\tilde h_1(k,t)=\int_{-\infty}^\infty h_1(y,t)\exp(-iky)\,\textrm{d}y.
\end{equation}
This concludes the derivation of the equation of contact line motion. The stationary version of
Eq.~(\ref{h1F}) was derived first by Pomeau and Vannimenus \cite{PomVan}.

The static contact line position is defined by the minimum of the energy $U[h(y)]$. However, it can exhibit
multiple minima that correspond to the metastable states. In this case one needs to choose between them
\cite{Garoff} to determine correctly the contact line position and corresponding local contact angle (which
are both different in advancing and receding cases). In dynamics, Eq.~(\ref{hh}) provides us with an unique
solution because the system ``knows" the direction of the contact line motion and its history.

\section{Simple example} To show the relevance of this approach, we solve rigorously the contact line dynamics for a simple case of a single
stripe-shaped defect at the wall,
\begin{equation}\label{cex}
  c(y,z)=\left\{\begin{array}{cc}
    c_d, & |y|\le \Delta, \\
    c_s, & |y|> \Delta,
  \end{array}\right.
\end{equation}
where $c_d,c_s\le 1$ and $\Delta$ are constants. Consider first the contact line displacement at equilibrium
described by the equations
\begin{equation}\label{eqil}
    h_0^{eq}=c_0l_c, \quad
    \tilde h_1^{eq}=\tilde c_1/\sqrt{l_c^{-2}+k^2}.
\end{equation}
Notice that $c_0$ should be small enough so that $f$ defined by Eq.~(\ref{f}) satisfies conditions
(\ref{weak}). Obviously, Eq.~(\ref{h0}) results in $c_0=c_s$ and
\begin{equation}\label{h1eq}
  \tilde h_1^{eq}={2\delta c\sin k\Delta\over k\sqrt{l_c^{-2}+k^2}},
\end{equation}
where $\delta c=c_d-c_s$.

To show how this result relates to earlier approaches to the single-defect problem, we analyze first the
limit $\Delta\rightarrow 0$.  The inverse transform can easily be taken using the tables \cite{Bateman} and
results in
\begin{equation}\label{hd}
h_1^{eq}(y)={2\delta c\Delta\over\pi}K_0(y/l_c),
\end{equation}
\begin{figure}[htb]
  \begin{center}
  \includegraphics[height=7cm]{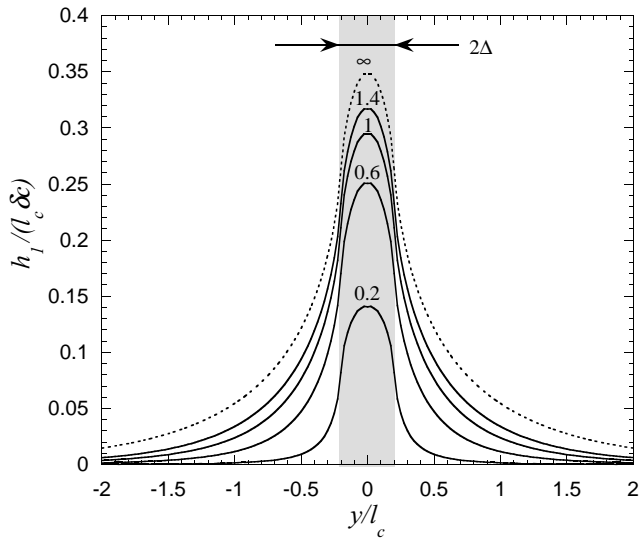}
  \end{center}
\caption{Time evolution of the deviation $h_1(y)$ of the contact line from its horizontally averaged value.
The defect area (half-width of which is $\Delta=0.2l_c$) is shadowed. The time values in the units $\xi
l_c/\sigma$ are shown above the corresponding curve.} \label{Stripe}
\end{figure}
where $K_0(\cdot)$ is the modified Bessel function of zeroth order. One recognizes the result \cite{JoRob}.
It leads to an unphysical divergence when $y\rightarrow 0$: $h_1^{eq}\sim -\log(y)$ and a cut-off at small
$y$ is needed \cite{DeGennes}.

Using the present approach, one can understand the origin of this divergence. The limit $\Delta\rightarrow 0$
in the rigorous result (\ref{h1eq}) is equivalent to the limit $k\rightarrow 0$ since their product enters
Eq.~(\ref{h1eq}). The asymptotics at $k\rightarrow 0$ in the Fourier transform corresponds  \cite{Korn} to
the limit $y\rightarrow\infty$ in the original function, which means that Eq.~(\ref{hd}) is correct only at
$y\rightarrow\infty$. However, it is not guaranteed that its asymptotics at $y\rightarrow 0$ is correct.

Let us return now to the rigorous expression (\ref{h1eq}). The  Fourier transform (\ref{h1eq}) can be
inverted analytically,
\begin{equation}\label{h1or}
h_1^{eq}(y)=\delta c[F(\Delta+y)+F(\Delta-y)]/\pi,
\end{equation}
where $F(y)=\int_0^yK_0(y/l_c\,)\,\textrm{d}y$. It is clear now that $h_1^{eq}(y\rightarrow 0)$ remains
finite, see the dotted curve in Fig.~\ref{Stripe}.

The dynamic solution can be obtained the same way by using the initial condition
$h(t=0)=0$:
\begin{equation}\label{h1d}
  \begin{array}{r}
    \displaystyle h_0(t)=c_sl_c\{1-\exp[-t\sigma/(\xi l_c)]\}, \\
    h_1(y,t)=\delta c\left[F(\Delta+y)+F(\Delta-y)\right.\left.-G(\Delta+y,t)-G(\Delta-y,t)\right]/\pi,
  \end{array}
\end{equation}
where $$G(y,t)=\int_0^yK_0\left(\sqrt{l_c^{-2}(y^2+t^2\sigma^2\xi^{-2})}\,\right)\textrm{d}y.$$ The time
evolution of $h_1(y)$ is shown in Fig.~\ref{Stripe}. These curves can be compared to those obtained
experimentally in \cite{Nad,Caz,Limat} where the motion of the contact line over a single defect was studied.
The comparison shows a good qualitative agreement. The results cannot be compared quantitatively since in
each of these articles some parameters that enter Eq.~\ref{h1d} ($\delta c$ in particular) are missing.

\section{Conclusion} Eq.~(\ref{hh}) describes the spontaneous contact line motion for
\emph{arbitrary} distribution of surface energy given by the function $c(y,z)$ (provided $|c|$ is small
enough so that the conditions (\ref{weak}) are satisfied). This function can be considered random and the
equation becomes stochastic. It can be used to establish any statistical parameter of the contact line in
dynamics. In particular, the collective effect of defects on the contact line motion can be studied.

In this work we solve a simple example where the defect properties do not vary along the average contact line
velocity vector. In the general case, where the local value of the equilibrium contact angle (or of the
function $c$) varies along this direction, the equation of the contact line motion becomes non-linear. More
sophisticated methods are then needed to solve it.

\acknowledgements We thank Y. Pomeau for the critical reading of this Letter.

\end{document}